
\headline={\ifnum\pageno=1\firstheadline\else
\ifodd\pageno\rightheadline \else\leftheadline\fi\fi}
\def\firstheadline{\hfil}
\def\rightheadline{\hfil}
\def\leftheadline{\hfil}
\footline={\ifnum\pageno=1\firstfootline\else\otherfootline\fi}
\def\firstfootline{\rm\hss\folio\hss}
\def\otherfootline{\hfil}

\font\tenrm=cmr10
\font\tenit=cmti10
\font\elevenbf=cmbx10 scaled\magstep 1
\font\elevenrm=cmr10 scaled\magstep 1
\font\elevenit=cmti10 scaled\magstep 1

\font\ninerm=cmr9

\nopagenumbers
\line{\hfil }
\hsize=6.0truein
\vsize=8.5truein
\parindent=3pc
\hfill UCLA/93/TEP/21
\vglue 1cm
\baselineskip=10pt
\centerline{\elevenbf CONNECTION ON THE THEORY SPACE\footnote{*}
{\ninerm\baselineskip=11pt Based on a talk given at
Strings '93 held during 24 -- 29 May in Berkeley.
This work was supported in part by the
D.O.E. contract DE-AT03-88ER 40384 Mod A006 Task C.\hfill}}
\vglue 1.0cm
\centerline{\tenrm HIDENORI SONODA}
\baselineskip=13pt
\centerline{\tenit Department of Physics, UCLA}
\baselineskip=12pt
\centerline{\tenit Los Angeles, CA 90024--1547, USA}
\vglue 0.8cm
\def\dl{{d \over dl}~}
\def\e{{\rm e}}
\def\O{{\cal O}}
\def\vev#1{\langle#1\rangle_g}
\def\ep{\epsilon}
\line{\elevenbf 1. Introduction \hfil}
\vglue 0.4cm
\baselineskip=14pt
\elevenrm
The idea of the theory space is essential for qualitative
understanding
of quantum field theory${}^1$; only through the notions of relevance,
irrelevance, and universality, we understand the meaning
of the continuum limit of a field theory.
But our use of the
theory space has so far been mostly qualitative.
It is the purpose of this short article to treat
the theory space seriously and introduce a
quantitative structure.$^{2-4}$

\vglue 0.6cm
\line{\elevenbf 2. Covariant Quantities in the Theory Space \hfil}
\vglue 0.4cm
We consider renormalized field theories in $D$ dimensional euclidean
space
parametrized by $g^i (i=1, ... ,N)$.  The parameters $g^i$ give local
coordinates
of the theory space.
We take the origin $g^i = 0 (\forall i)$ to be
the UV fixed point.  The renormalization group (RG)
describes the change of the theory under scale change;
under the transformation $R_l$ the renormalization scale
is fixed, but the coordinate distance $r$ transforms into
$r \e^{- l}$.  The RG equations of the
parameters are given by
$$
\dl g^i = \beta^i (g) \equiv x_i g^i + {\rm O} \left( g^2 \right) ,
\eqno (2.1)
$$
where the scale dimension $x_i$ of the parameter $g^i$ is positive.
Note that the beta function $\beta^i (g)$ is a vector field on the
theory space.

Each point in the theory space is a renormalized field theory which
has
an infinite number of linearly independent composite fields.  Let
$\{\Phi_a\}_g$ be a complete basis of composite fields at $g$.  The
space
of composite fields make a linear space, and therefore the composite
fields make an infinite dimensional vector bundle over the theory
space.
The choice of the local basis is by no means unique, and two bases
$\{\Phi_a\}_g$ and $\{\Phi'_a\}_g$ are related by an invertible
infinite
dimensional matrix $N (g)$ as
$$
\Phi'_a = (N(g))_a^{~b} \Phi_b .\eqno (2.2)
$$
Therefore, the correlation function of $n$ composite fields
$$
\vev{\Phi_{a_1} (r_1) ... \Phi_{a_n} (r_n)} \eqno (2.3)
$$
forms a rank-n tensor field on the theory space.

We can write the RG equations of the composite fields as
$$
\dl \Phi_a = \left( \Gamma (g)\right)_a^{~b} \Phi_b , \eqno (2.4)
$$
where the matrix
$$
\left(\Gamma (g)\right)_a^{~b} = y_a \delta_a^b + {\rm O} (g) \eqno
(2.5)
$$
gives the scale dimensions $y_a$ and anomalous dimensions
of the renormalized fields.  The precise meaning of the above RG
equations
is given by the following RG equations for correlation functions:
$$
\eqalignno{
&\dl \vev{\Phi_{a_1} (r_1) ... \Phi_{a_n} (r_n)}
\equiv \left( - \sum_{k=1}^n r_{k,\mu} {\partial \over \partial
r_{k,\mu}}
+ \beta^i {\partial \over \partial g^i} \right) \vev{\Phi_{a_1} (r_1)
... \Phi_{a_n} (r_n)} \cr
&\quad = \sum_{k=1}^n \Gamma_{a_k}^{~~b} ~\vev{\Phi_{a_1} (r_1) ...
\Phi_b (r_k) ... \Phi_{a_n} (r_n)} .& (2.6)\cr}
$$

Among the infinite number of composite fields,
$N$ composite scalar fields stand out; they are the fields
$\O_i$ {\it conjugate} to the parameters $g^i$.  The notion
of conjugacy should be familiar from thermodynamics,
where the external magnetic field $H$ has magnetization
$M$ as its conjugate field.  The expectation value
of $\O_i$ gives the derivative of the free energy
density $F(g)$ with respect to $g^i$:
$$
\vev{\O_i} = {\partial \over \partial g^i}~F(g) .\eqno (2.7)
$$
Since the free energy density satisfies the canonical RG equation
$$
\dl F(g) = D F(g) \eqno (2.8)
$$
($D$ is the dimensionality of space), we find, from Eqs.~(2.1) and
(2.8),
that
$$
\dl \O_i = D \O_i - {\partial \beta^j \over \partial g^i}~\O_j .\eqno
(2.9)
$$
We will define the conjugate fields more precisely in sect.~3.
\vglue 0.6cm
\line{\elevenbf 3. Variational Formula\hfil}
\vglue 0.4cm
We now consider the $g$-dependence of the correlation functions,
which are tensor fields on the theory space.  In analogy to
thermodynamics
and statistical mechanics, the derivative of a correlation function
with respect to $g^i$ should be given as a volume integral of the
conjugate field $\O_i$:
$$
\eqalignno{&- {\partial \over \partial g^i} \vev{\Phi_{a_1} (r_1) ...
\Phi_{a_n} (r_n)} \cr
&\quad = \lim_{\ep \to 0} \Bigg[ \int_{|r-r_k| \ge \ep}
d^D r~\vev{\left(\O_i (r) - \vev{\O_i}\right)
\Phi_{a_1} (r_1) ... \Phi_{a_n} (r_n)} \cr
&\quad\quad + \sum_{k=1}^n \left( {c_{i,a_k}}^b (g) - \int_{1 \ge r
\ge \ep}
{d^D r \over {\rm vol} (S^{D-1})}~ {C_{i,a_k}}^b (r;g) \right) \vev{
\Phi_{a_1} (r_1) ... \Phi_b (r_k) ... \Phi_{a_n} (r_n)} \Bigg] ,&
(3.1)\cr}
$$
where $\rm{vol} (S^{D-1})$ is the volume of a unit $D$--$1$ sphere.
Two remarks are in order.  First, due to the short distance
singularity
$$
\O_i (r) \Phi_a (0) = {1 \over {\rm vol} (S^{D-1})}~{C_{i,a}}^b (r;g)
\Phi_b +
{\rm O} \left( {1 \over r^D} \right) ,\eqno (3.2)
$$
the integral needs subtractions.
In the above operator product expansion (OPE), we
keep only composite fields $\Phi_b$ whose scale dimension is at most
$y_a$, since we need to subtract only the unintegrable part.
Second, we need the finite counterterms ${c_{i,a}}^b (g)$ not only
to compensate the arbitrariness of the subtraction scheme,
but also to satisfy covariance in the theory space.  Under a change
of
basis, Eq.~(2.2), the subtracted integral on the right-hand side
of Eq.~(3.1) is covariant, but the left-hand side is non-covariant
due to the naked derivative.  Hence, covariance demands the
finite counterterms ${c_{i,a}}^b (g)$ which transform as a connection
$$
c_i (g) \to c'_i (g) = N(g) \left( \partial_i + c_i (g) \right)
N^{-1} (g)
\eqno (3.3)
$$
under the change of basis Eq.~(2.2).  This is how the connection
$c_i$ arises naturally in the theory space.  We call Eq.~(3.1)
as a variational formula.
\vglue 0.6cm
\line{\elevenbf 4. Consistency Conditions \hfil}
\vglue 0.4cm
It is difficult to {\elevenit derive} the variational formula
from first principles.  Instead, we {\elevenit demand} the conjugate
fields $\O_i$ to satisfy Eq.~(2.7) and the variational formula
Eq.~(3.1).
It is an assumption that such fields exist.  Nonetheless, we can
check the consistency of Eq.~(3.1).

First, we demand that Eq.~(3.1) be consistent with the RG equations
(2.1), (2.4), and (2.9).  Then, we obtain
$$
C_i (r=1;g) = {\partial \over \partial g^i}~\Psi (g) + [ c_i (g) ,
\Psi (g) ]
+ \beta^j (g) \Omega_{ji} (g) ,\eqno (4.1)
$$
where
$$
\Psi (g) \equiv \Gamma (g) + \beta^i (g) c_i (g) \eqno (4.2)
$$
is a covariant tensor field, and the curvature $\Omega_{ij}$ is
defined by
$$
\Omega_{ij} \equiv \partial_i c_j  - \partial_j c_i + [c_i, c_j]
.\eqno (4.3)
$$
Eq.~(4.1) gives a geometric expression of the singular part of
the OPE coefficients.

Second, we demand that Eq.~(3.1) satisfy the Maxwell relation:
$$
\partial_i \partial_j = \partial_j \partial_i .\eqno (4.4)
$$
This assures the consistency of using Eq.~(3.1) recursively
to evaluate higher order derivatives.  Eq.~(4.4) gives the curvature
in terms of a double integral over a finite domain:
$$
\Omega_{ij} ~\vev{\Phi} = \int_{r \le 1} d^D r ~{\rm F. P.} \int_{r'
\le 1} d^D r'~
\vev{\O_j (r) \left( \O_i (r') - {C_i (r') \over {\rm vol}(S^{D-1})}
\right) \Phi (0)
- ( i \leftrightarrow j ) } ,\eqno (4.5)
$$
where F.P. denotes taking the integrable part with respect to $r$.
\vglue 0.6cm
\line{\elevenbf 5. Conclusions \hfil}
\vglue 0.4cm
By studying the variational formula, Eq.~(3.1), we have found that
the geometrical quantities on the theory space, i.e.,
the vector field $\beta^i (g)$ (beta functions), the
rank-two tensor field ${\Psi_a}^b (g)$ (scale and anomalous
dimensions),
and the connection $c_i$ (finite counterterms), determine
the short-distance singularities of the theory, as is given by
Eqs.~(4.1) and (4.5).  These are analogues of the well-known
relation between the renormalization constants and the beta functions
and anomalous dimensions in the minimal subtraction scheme.$^5$
It is important to note that
the connection $c_i$ has more than its geometric significance:
the connection gives the finite counterterms in the variational
formula (3.1).

One promising place where we can apply the variational formula
and the connection $c_i$ of sect.~3 is in closed string field theory.
To understand the background independence of closed
string field theory, we must find the
relation between two different formulations
based upon two conformal field theories.  This requires comparing
string states (or vertex operators) in one conformal
field theory to those in the other conformal
field theory.  Hence, we
need a connection to parallel transport states from one theory
to the other.  We have applied the variational formula of sect.~3 to
the space
of conformal field theories$^{6,7}$; in particular
we have studied the properties of the connection $c_i$ and its two
modifications.
 It is not yet clear which precise connection is the right
choice
for closed string field theory, but a connection,
together with the variational formula (3.1), is expected to play an
important role.$^8$

\vglue 0.6cm
\line{\elevenbf References \hfil}
\vglue 0.4cm
\item{1.} K. Wilson and J. Kogut, {\elevenit Phys. Rep.} {\elevenbf
C12} (1974) 75.
\item{2.} H. Sonoda, {\elevenit Nucl. Phys.}
{\elevenbf B383} (1992) 173.
\item{3.} H. Sonoda, {\elevenit Nucl. Phys.}
{\elevenbf B394} (1993) 302.
\item{4.} H. Sonoda, {\elevenit Current-Current Singularities under
External Gauge Fields} (hep-th 9212025).
\item{5.} G. 't Hooft, {\elevenit Nucl. Phys.} {\elevenbf B62} (1973)
444.
\item{6.} K. Ranganathan, H. Sonoda, and B. Zwiebach, {\elevenit
Connections on the State-Space over Conformal Field Theories}
(hep-th 9304053).
\item{7.} B. Zwiebach, contribution to this proceedings.
\item{8.} A. Sen and B. Zwiebach, work in progress.

\vfill
\bye